\documentclass[english,12pt,a4paper]{article}
\usepackage{authblk}
\usepackage[text={17.0cm,23.7cm}]{geometry}
\usepackage[T1]{fontenc}
\usepackage[utf8]{inputenc}
\usepackage{babel}
\usepackage{hyperref}
\usepackage{cite}
\usepackage{graphicx}
\usepackage{amsmath}
\usepackage{amssymb}
\usepackage[dvipsnames]{xcolor}
\usepackage{yfonts}
\usepackage{ulem}
\usepackage{enumitem}
\usepackage{slashed}
\usepackage{textcomp}
\usepackage{subcaption}
\usepackage{multirow}%fancy table

\hypersetup{colorlinks,citecolor=blue,linkcolor=blue,urlcolor=blue}
\newcommand{\second}[0]{\,\mathrm{s}}

\newcommand{\sigmav}[0]{\ensuremath{\langle \sigma v \rangle}}

\newcommand{\Ztwo}[0]{\ensuremath{\mathbb{Z}_2}}

\newcommand{\eV}[0]{\,\mathrm{eV}}
\newcommand{\keV}[0]{\,\mathrm{keV}}
\newcommand{\MeV}[0]{\,\mathrm{MeV}}
\newcommand{\GeV}[0]{\,\mathrm{GeV}}

\newcommand{\heff}{h_\mathrm{eff}}

\newcommand{\micromegas}{\textsc{micrOMEGAs}}

\begin{document}
	\title{\vspace{-2cm}
		{\normalsize
			\flushright TUM-HEP 1321/21\\}
		\vspace{0.6cm}
		\textbf{Production and signatures of multi-flavour dark matter scenarios with $t$-channel mediators}\\[8mm]}
	\author[a]{Johannes Herms}
	\author[b]{Alejandro Ibarra}
	\affil[a]{\normalsize\textit{Max-Planck-Institut f\"ur Kernphysik, Saupfercheckweg 1, 69117 Heidelberg, Germany}}
	\affil[b]{\normalsize\textit{Physik-Department, Technische Universit\"at M\"unchen, \protect\\James-Franck-Stra\ss{}e, 85748 Garching, Germany}}
	\date{}

\maketitle
\begin{abstract}
We investigate the phenomenology of a dark matter scenario containing two generations of the dark matter particle, differing only by their mass and their couplings to the other particles, akin to the quark and lepton sectors of the Standard Model. For concreteness, we consider the case where the two dark matter generations are Majorana fermions that couple to a right-handed lepton and a scalar mediator through Yukawa couplings. We identify different production regimes in the multi-flavor dark matter scenario and we argue that in some parts of the parameter space the heavier generation can play a pivotal role in generating the correct dark matter abundance. In these regions, the strength of the dark matter coupling to the Standard Model can be much larger than in the single-flavored dark matter scenario.  Correspondingly the indirect and direct detection signals can be significantly boosted. We also comment on the signatures of the model from the decay of the heavier dark matter generation into the lighter.
\end{abstract}

\section{Introduction}
\label{sec:intro}
Identifying the nature of the Dark Matter (DM) in the Universe is a central goal of contemporary cosmology and Particle Physics. Among the many candidates proposed in the literature, dark matter in the form of thermal relic particles is of particular interest, since it links the observed dark matter density via its production to  signatures that can be searched for at dedicated experiments at the Earth or in space (for reviews, see {\it e.g.} \cite{Jungman:1995df,Bertone:2004pz,Bergstrom:2000pn,Feng:2010gw}).

The Standard Model of particle physics (SM) needs to be extended to accommodate a viable dark matter candidate. Minimal extensions consist of a single dark matter particle and a portal interaction to the visible sector. 
On the other hand, it is a peculiar feature of Nature that all Standard Model fermions appear in three flavours,  sharing all gauge quantum numbers and differing only by their coupling to the Higgs field, and correspondingly by their mass.
Current astronomical and cosmological observations are largely insensitive to the heavier quark and charged lepton generations.\footnote{Observations of the cosmic microwave background radiation~\cite{Aghanim:2018eyx} and of the observed abundances of primordial elements~\cite{Steigman:2012ve} point towards three generations of active neutrinos.} However, the heavier generations could have played an important role in the very early stages of the Universe, for example for baryogenesis, and are crucial for the correct interpretation of laboratory experiments. 

In this paper we entertain the possibility that Nature could contain several generations of the dark matter particle, akin to the fermion sector of the Standard Model. This may seem an unnecessary complication, since current astronomical and cosmological data can be well explained by a single dark matter component. However, and mirroring the well known conclusions for the visible sector, we will argue that the heavier dark matter generations could have played a role in the early Universe, and that they could leave novel signatures in dark matter search experiments. 

For concreteness, we will consider the case where the two dark matter generations are Majorana fermions that couple to a right-handed lepton and a scalar mediator through Yukawa couplings. However, a similar rationale could be applied to other multi-flavor dark matter scenarios, with analogous conclusions. We will present our model in Section~\ref{sec:model}. In Section~\ref{sec:fermionWIMP} we will investigate the impact of the second dark matter flavour on the dark matter relic abundance. In Section~\ref{sec:leptoSignatures} we will discuss possible signatures of the heavier generation in cosmological observations, as well as in direct, indirect and collider experiments searching for dark matter. Finally, we will present our conclusions in Section~\ref{sec:conclusion}.

\section{Toy model}
\label{sec:model}

We consider a dark sector containing two generations of a Majorana fermion, $\psi_i$, $i=1,2$, singlets under the Standard Model gauge group, and odd under a global $\mathbb{Z}_2$  symmetry. The two Majorana fermions $\psi_1$ and $\psi_2$ only differ by their masses, $m_1$ and $m_2$, and possibly by their couplings to other particles (mirroring the quark and lepton sectors of the Standard Model). The $\mathbb{Z}_2$ symmetry is assumed to be exact in the electroweak vacuum and distinguishes between visible sector particles and hidden sector particles;  the former are assumed to be even under the $\mathbb{Z}_2$ symmetry and the latter odd.  As a result of this symmetry, the lightest particle of the hidden sector is absolutely stable and becomes a dark matter component, although there might be additional cosmologically long-lived particles in the hidden sector. We also assume the existence of a  $\mathbb{Z}_2$-odd scalar, $\Sigma$, with mass $m_\Sigma>m_2>m_1$ and gauge quantum numbers that allow a Yukawa coupling to $\psi_i$ and to some Standard Model fermion $f_\alpha$.

The Lagrangian of the model contains the following terms
\begin{align}
\label{eqn:LagrangianLeptophilic}
{\cal L}\supset
\left(\mathcal{D}_\mu \Sigma \right)^\dagger \left(\mathcal{D}^\mu \Sigma \right)
- m_\Sigma \Sigma^\dagger \Sigma 
- \lambda_{H\Sigma} |H|^2 |\Sigma|^2
+\Big( \frac{1}{2} \overline{\psi_i} i\slashed \partial \psi_i
- \frac{1}{2} m_i \overline{\psi_i^c} \psi_{i}
- g_{\alpha i} \overline{f_\alpha}  \psi_i \Sigma + \mathrm{h.c.}\Big)\;.
\end{align}
In order to simplify our analysis, and in order to highlight the new features of the model, we will  neglect the Higgs portal interaction to the hidden sector, $\lambda_{H\Sigma}=0$, and assume that the hidden-sector particles couple to a single generation of the fermion $f_\alpha$.  The coupling to different generations of Standard Model fermions would have implications for flavor physics in the visible sector and could provide complementary information about the characteristics of the dark sector; that analysis is however beyond the scope of this paper. We will further assume that $f$ is a right-handed lepton $l=e,\mu,\tau$.\footnote{The case where $f$ is a left-handed doublet leads  to neutrino masses at one-loop order and has received a lot of attention; this is the so-called scotogenic model~\cite{Ma:2006km}. The simultaneous explanation of dark matter and neutrino masses in this model requires multiple flavours of dark sector fermions and consequences for dark matter phenomenology have been explored by~\cite{Molinaro:2014lfa,Hessler:2016kwm}.}
After specifying  the lepton generation $\Sigma$ couples to,
the two-flavour model is then specified by three masses and two couplings:
$$ m_\Sigma, \; m_2, \; m_1 , \; g_2 , \; g_1 \,.$$

With this set-up, $\psi_1$ is the lightest $\mathbb{Z}_2$-odd particle and hence absolutely stable, while $\psi_2$ can decay into $\psi_1$ and visible sector particles through a virtual $\Sigma$. The dominant decay channel is $\psi_2\rightarrow \psi_1 l^+ l^-$. The decay width in the limit  $m_l \ll m_{\psi_2} \ll m_\Sigma$ reads~\cite{Garny:2010eg}
\begin{equation} %(2.3)
 \label{eqn:fermionDecayRateThreeBody}
 \Gamma_{\psi_2\to \psi_1 l^+l^-} = 
 \frac{|g_{1} g_{2}|^2}{2^{10} \pi^3 3}
 \frac{m_{\psi_2}^5}{m_\Sigma^4}
 \left( F_1(m_{\psi_1}^2/m_{\psi_2}^2)
 + 2 F_2(m_{\psi_1}^2/m_{\psi_2}^2)
 \right)
 \,,
\end{equation}
with
\begin{align}
 F_1(x) &= (1-x^2)(1-8x+x^2) - 12 x^2 \ln(x) \;, \\
 F_2(x) &= \sqrt{x} \left[ (1-x)(1+10x+x^2) + 6x(1+x)\ln(x) \right]  \;.
\end{align}
The decay rate is proportional to $\left|g_{1} g_{2}\right|^2$, and can range over many orders of magnitude.
Additional decay channels appear at the one loop level, for example $\psi_2\rightarrow\psi_1\gamma$, $\psi_2\rightarrow \psi_1 \nu\bar \nu$, or $\psi_2\rightarrow \psi_1 \pi^0, Z, h$ (when kinematically allowed).
Although these channels have a smaller width, they can be relevant for the experimental searches of the model. Among these channels, $\psi_2\rightarrow\psi_1\gamma$ is of particular significance, as it leads to a gamma-ray with fixed energy
\begin{equation}
 E_\gamma = \frac{m_{\psi_2}}{2} \left( 1-\frac{m_{\psi_1}^2}{m_{\psi_2}^2} \right)\,,
\end{equation}
which results in a monochromatic signal, which is routinely searched for in the gamma ray sky (e.g.~\cite{Essig:2013goa,Ackermann:2015lka}).
This allows for tests of the multiflavor dark matter scenario, as we will discuss in the next Section. The width of this channel, also in the limit $m_{\psi_2} \ll m_\Sigma$,
reads~\cite{Garny:2010eg}: 
\begin{equation}
 \label{eqn:fermionDecayRateLowMassLimit}
 \Gamma_{\psi_2\to\psi_1\gamma} = \frac{e^2 |g_{1} g_{2}|^2}{2^{15} \pi^5}\frac{m_{\psi_2}^5}{m_\Sigma^4} \left(1-\frac{m_1^2}{m_2^2} \right)^3 \left(1-\frac{m_1}{m_2} \right)^2\,.
\end{equation}

The charged scalar $\Sigma$ is expected to be in equilibrium with the SM bath in the early Universe and is the portal connecting the dark and visible sectors. Concretely, it mediates the production/annihilation process $\psi_i\psi_j\leftrightarrow l^+l^-$, as well as the conversion process $\psi_i l \leftrightarrow \psi_j l$.
If $\Sigma$ is close in mass to $\psi_i$, $\Sigma\Sigma$ pair annihilations or $\Sigma \psi_i$ coannihilations can be important to determine the $\psi_i$ abundance.
In the case where the coupling $g_i$ is very small, dark matter production can be dominated by $\Sigma$ decay via freeze-in.

The lifetime of the charged scalar is much shorter than the age of the Universe, since the decay rate is only proportional to $\left|g_{i}\right|^2$.
The existence of this new charged particle in the particle spectrum can lead, however,  to signals at colliders. If $\Sigma$ is sufficiently long-lived, it is expected to leave a highly ionizing charged track in a detector. The non-observation of such events at the LHC sets the lower limit $m_\Sigma > 430 \GeV$~\cite{Aaboud:2019trc}. In contrast, if $\Sigma$ is short lived, it decays into a charged lepton and a dark matter particle, which goes undetected. In this regime, the limits on the mass are somewhat weaker, and depend on the mass difference between $\Sigma$ and the dark matter particle ~\cite{Sirunyan:2018nwe,Aaboud:2018jiw,Aad:2019byo,Aaboud:2019trc}. In some scenarios the mediator $\Sigma$ is close in mass to the dark matter particle. As a result, the charged lepton is too soft and the decay products are all undetected. Yet, the production of $\Sigma$ may be accompanied by radiation from the initial state, which could be detected. The null search results for an excess of initial state radiation and missing transverse energy  at the LHC sets a model dependent limit on the mass of the charged scalar particles, which can be as large as $m_\Sigma > 190\GeV$~\cite{Aaboud:2017leg}.
In most of our analysis we will adopt $m_\Sigma=430$ GeV as benchmark value for the mass of the charged mediator, while for a representative near-degenerate benchmark we will adopt $m_\Sigma=200$ GeV.

\section{General relic abundances from FIMP to WIMP}
\label{sec:fermionWIMP}
The evolution of the abundances of $\Sigma$ and $\psi_{1,2}$ in the early Universe is determined by three coupled Boltzmann equations.
Assuming kinetic equilibrium of all species involved,\footnote{
The validity of this assumption in a related single-component DM model has been studied by~\cite{Garny:2017rxs},
who find that the impact of deviations from kinetic equilibrium on the relic abundance is small
(see however~\cite{DAgnolo:2017dbv}).
}
the comoving number densities $n_{\Sigma,\psi_2,\psi_1}$ obey:
\begin{equation}
\label{eqn:boltzmannN}
 \begin{split}
 \frac{d n_a}{dt} + 3 H n_a =
 &- \sum_{b} \sigmav_{ab \to AB}^\mathrm{ann} \left( n_a n_b - n_a^\mathrm{eq} n_b^\mathrm{eq} \right) \\
 &- \sum_{b} \sigmav_{aA \to bB}^\mathrm{sca} \left( n_a n_A^\mathrm{eq} - n_b n_B^\mathrm{eq} \frac{n_a^\mathrm{eq}}{n_b^\mathrm{eq}} \right) \\
 &- \sum_{b} \tilde \Gamma_{a \to b} \left( n_a - n_b \frac{n_a^\mathrm{eq}}{n_b^\mathrm{eq}} \right)\,,
 \end{split}
\end{equation}
where $a,b$ refer to particle species in the dark sector while $A,B$ refer to Standard Model particles, assumed to form an equilibrium thermal bath.
$H$ is the Hubble expansion rate of the Universe.
All particle distribution functions are approximated by Maxwell-Boltzmann distributions.
The $\sigmav_{ab \to AB}^\mathrm{ann}$ term describes annihilations of \Ztwo-odd particles into Standard Model bath particles, while the $\sigmav_{aA \to bB}^\mathrm{sca}$ and $\tilde \Gamma_{a \to b}$ terms describe conversion processes between \Ztwo-odd particles.
Here, $\sigmav$ and $\tilde \Gamma$ are respectively the thermally averaged cross section and decay rate~\cite{Gondolo:1990dk}.
The processes relevant for the evolution of the number densities are listed in Table~\ref{tab:processes}, along with their parametric dependence on the Yukawa couplings $g_1$ and $g_2$. In our analysis we will neglect chirality-suppressed processes involving neutrinos, as well as processes suppressed by the small lepton Yukawa couplings. 

\begin{table}[t!]
 \begin{center}
    \begin{tabular}{|c c|c c|c|}
      \hline
      \multicolumn{2}{|c|}{initial state} & \multicolumn{2}{|c|}{final state} & $g_i$ scaling\\
      \hline
      \hline
      \multirow{8}{*}{$\Sigma^+$} & \multirow{8}{*}{$\Sigma^-$} & $\gamma$ & $\gamma$  & \multirow{7}{*}{$1$} \\
      \cline{3-4}
      &  & $\gamma$ & $Z$ & \\
      \cline{3-4}
      &  & $Z$ & $Z$ & \\
      \cline{3-4}
      &  & $W^+$ & $W^-$ & \\
      \cline{3-4}
      &  & $H$ & $H$ & \\
      \cline{3-4}
      &  & $H$ & $Z$ & \\
      \cline{3-4}
      &  & $q$ & $\bar q$ & \\
      \cline{3-5}
      &  & $l^+$ & $l^-$ & $1, g_1^2, g_1^4, g_2^2, g_2^4, g_1^2 g_2^2$ \\
      \hline
      $\Sigma^+$ & $\Sigma^+$ & $l^+$ & $l^+$  & $g_1^4, g_1^2g_2^2,g_2^4$ \\
      \hline
      $\psi_i$ & $\Sigma^+$ & $l^+$ & $\gamma,Z$  & $g_i^2$ \\
      \hline
      $\psi_i$ & $\psi_j$ & $l^+$ & $l^-$  & $g_i^2 g_j^2$ \\
      \hline
      \hline
      $\psi_i$ & $l^\pm$ & $\psi_j$ & $l^\pm$  & $g_i^2 g_j^2$ \\
      \hline
      $\Sigma^\pm$ & $\gamma,Z$ & $\psi_i$ & $l^\pm$  & $g_i^2$ \\
      \hline
       $\Sigma^\pm$ & $l^\pm$ & $\psi_i$ & $\gamma$  & $g_i^2$ \\
      \hline
      \hline
      \multicolumn{2}{|c|}{$\Sigma^\pm$} & $\psi_i$ & $l^\pm$ & $g_i^2$ \\
      \hline
      \multicolumn{2}{|c|}{$\psi_2$} & $\psi_1$ & $l^+ l^-$ & $g_1^2 g_2^2$ \\
      \hline
    \end{tabular}
  \end{center}
    \caption{Annihilation, scattering and decay processes included in the Boltzmann equation, along with their parametric dependence on the Yukawa couplings $g_1$ and $g_2$.
    }
    \label{tab:processes}
\end{table}

For the numerical solution of Eqn.~\eqref{eqn:boltzmannN} it is convenient to use as variables the abundances $Y_a=n_a/s$ with $s$ the entropy density, as functions of $x=m_1/T$ with $T$ the temperature of the SM bath:
\begin{equation}
\label{eqn:boltzmannLNY}
 \begin{split}
 \frac{d \ln Y_a}{dx} =
 &- \sum_{b} \left[\frac{\sigmav_{ab \to AB}^\mathrm{ann} s Y_a \frac{Y_b^\mathrm{eq}}{Y_a^\mathrm{eq}}}{x \tilde H}\right] \left( \frac{Y_b}{Y_a} \frac{Y_a^\mathrm{eq}}{Y_b^\mathrm{eq}}  - \frac{Y_a^{\mathrm{eq}\,2}}{Y_a^2} \right) \\
 &- \sum_{b} \left[ 
 \frac{\sigmav_{aA \to bB}^\mathrm{sca} s Y_A^\mathrm{eq}}{x \tilde H}
 %\frac{Y_i}{Y_i} %no need to write this
 + \frac{\tilde{\Gamma}_{a\to b}}{x \tilde H}
 \right]
 \left( 1 - \frac{Y_b}{Y_a} \frac{Y_a^\mathrm{eq}}{Y_b^\mathrm{eq}} \right)\,,
 \end{split}
\end{equation}
where $\tilde H = H/\tilde g$ with $\tilde g = 1+\frac{1}{3} \frac{T}{\heff} \frac{d\heff}{dT}$, and $\heff$ the effective number of degrees of freedom contributing to the entropy density at the temperature $T$. We have solved the Boltzmann equations numerically, adopting as initial condition $Y_a(x\to 0) = 0$.\footnote{
To calculate the thermally averaged cross sections we use \textsc{Feyn\-Rules}~\cite{Alloul:2013bka}, \textsc{Calc-HEP}~\cite{Belyaev:2012qa} and \textsc{micrOMEGAs}~\cite{Belanger:2018mqt}. We note that the current version of \micromegas\  cannot be applied to the full parameter space of our multiflavor dark matter scenario, but only when the conversion terms are either very large or very small compared to the Hubble rate.  In order to completely cover the parameter space of couplings, we have solved the set of Boltzmann equations numerically, and only used the public codes to calculate cross sections.}
Note that the conversion reaction $\Sigma \gamma \leftrightarrow \psi_i e$ and its crossed process have a soft divergence and a near $t$-channel divergence related to a nearly on-shell electron propagator. These are regulated by introducing thermal masses~\cite{Heeba:2019jho}
for the electron and photon in these reactions. The related process $\Sigma Z \leftrightarrow \psi_i e$ is treated in a similar manner, however we find numerically that this process is subdominant and can be safely neglected.  The  crossed process $\Sigma e \to \psi_i Z$ is $t$-channel divergent due to an on-shell electron propagator, which can be attributed to a double counting of the $\Sigma$ decay process. We expect this process to be subdominant compared to the analogous process involving a photon, and we have neglected it in our numerical calculation.

\begin{figure}[t!]
 \begin{center}
   \includegraphics[width=0.297\textwidth]{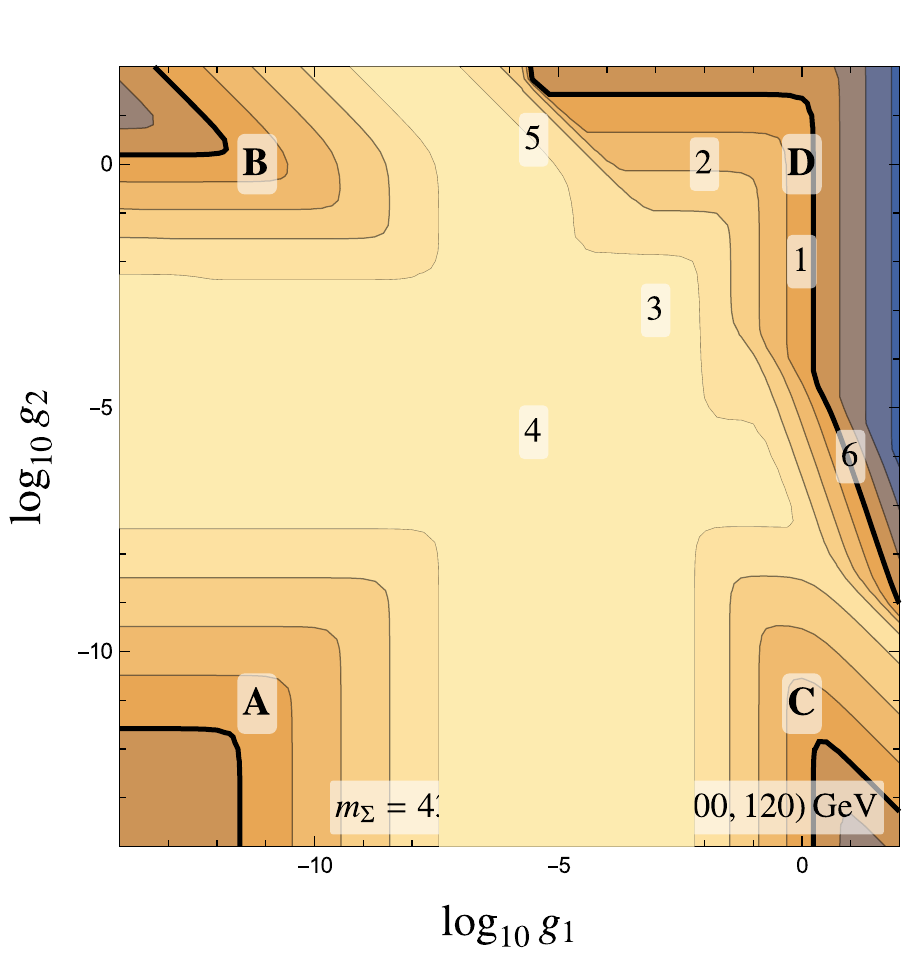}
   \includegraphics[width=0.297\textwidth]{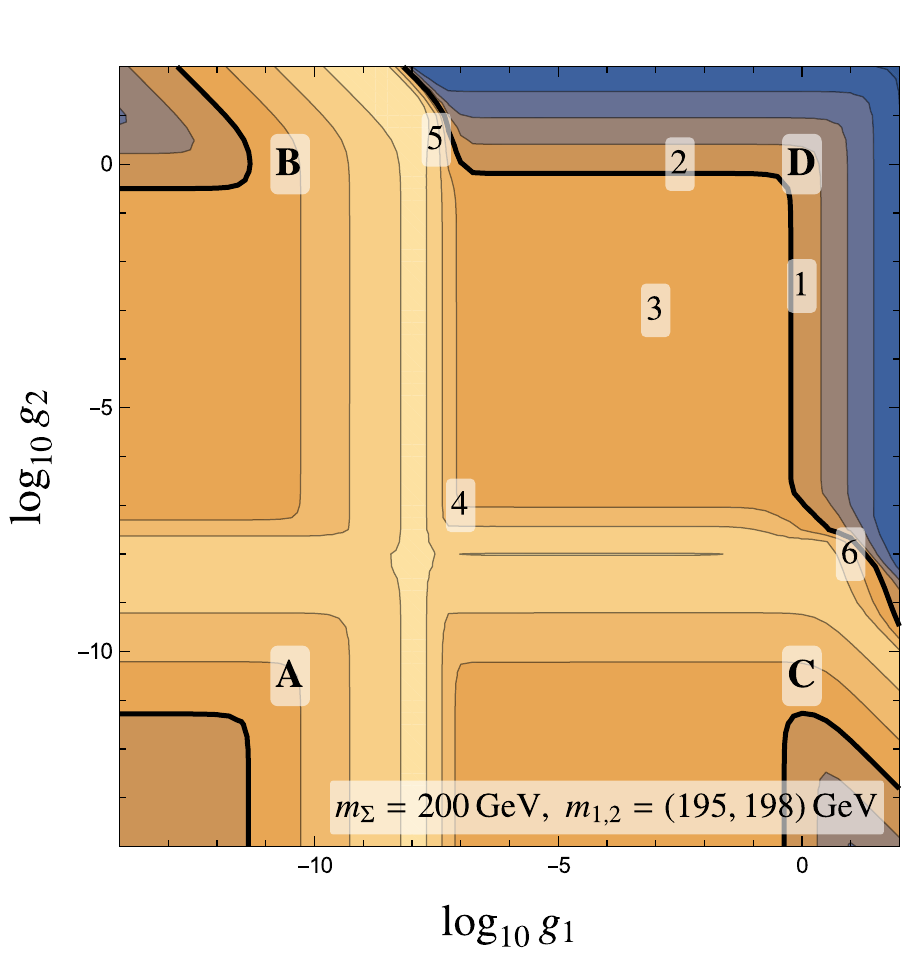}
   \includegraphics[width=0.357\textwidth]{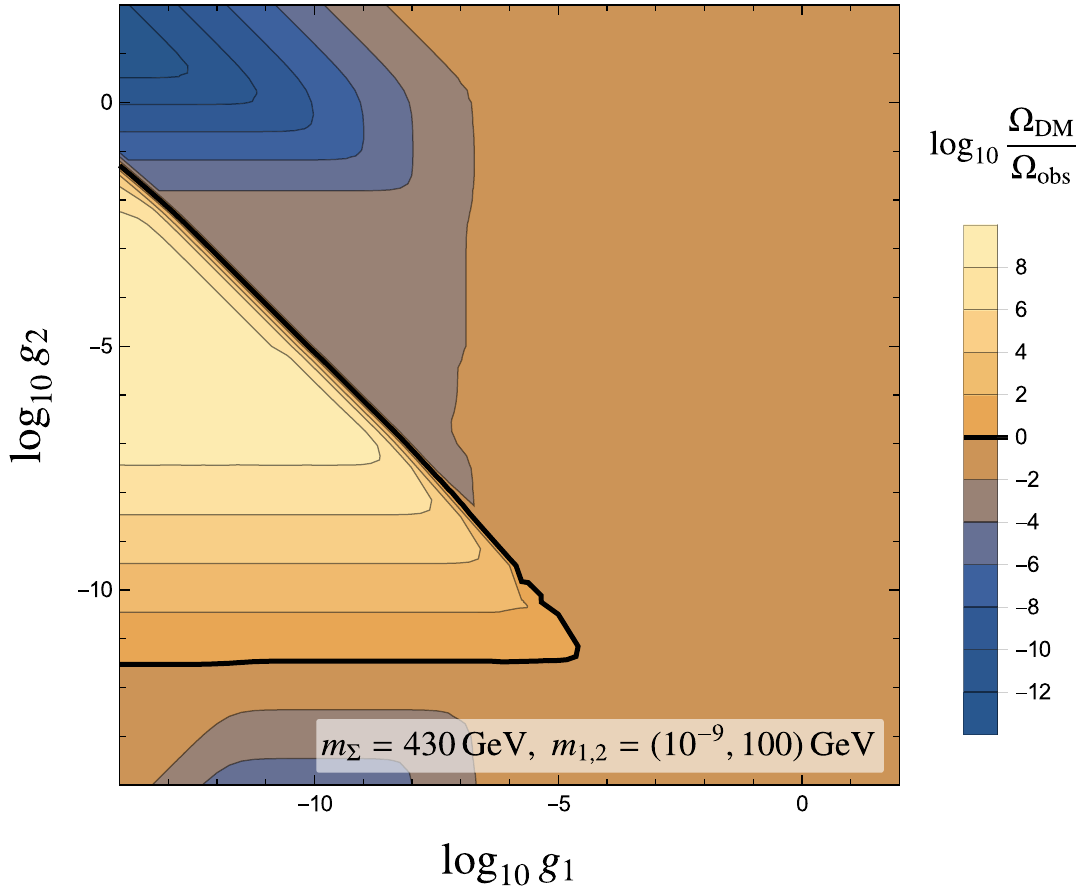}
 \end{center}
 \caption{
 Isocontours of the dark matter relic density predicted in our multi-flavor scenario relative to the observed dark matter density, in the parameter space spanned by the two Yukawa couplings  $g_1,g_2$, for the generic benchmark (left panel), the near-degenerate benchmark (middle panel) and the very hierarchical benchmark (right panel). The description of the labels can be found in the main text. 
 }
 \label{fig:relicContours}
\end{figure}

Figure~\ref{fig:relicContours} shows contours of the dark matter relic abundance compared to the value determined by CMB measurements~\cite{Aghanim:2018eyx} in the plane of couplings $g_{1,2}$ for three benchmark sets of parameters $(m_\Sigma,m_1,m_2) = (430,100,120)$ (left panel),  $(200,195,198)$ (middle panel) and $(430,10^{-9},100)$ GeV (right panel), chosen respectively as one generic benchmark, a near-degenerate benchmark, and one very hierarchical benchmark allowing for a hot dark matter component. In these figures we have left the couplings to vary between $10^{-14}$ and 100, to identify all relevant features of the parameter space. On should note that the validity of our perturbative calculation requires $g_i \lesssim \sqrt{4\pi}$. Therefore, in the analysis of the signals we will disregard the region with larger couplings, bearing in mind that it requires a separate analysis and is not necessary excluded by observations. 

The main qualitative features of our multiflavor dark matter scenario are already present in the generic benchmark, characterized by  $m_\Sigma=430$ GeV, $m_1=100$ GeV and $m_2=120$ GeV.
We find that most of the parameter space is ruled out, since the predicted contribution from $\psi_1$ (and possibly also $\psi_2$) to the dark matter abundance is many orders of magnitude larger than the observed value. Four distinct regions are compatible with observations, depending on whether the particle species $\psi_1$ and $\psi_2$ thermalise or not:
\begin{enumerate}[label=(\Alph*)]
 \item Neither $\psi_1$ nor $\psi_2$ thermalise, and their abundance is set by freeze-in.
 \item $\psi_1$ never thermalises, while $\psi_2$ does. The abundance of $\psi_1$ is set by freeze-in and the abundance of $\psi_2$ by freeze-out. 
 \item $\psi_1$  thermalises, while $\psi_2$ does not. The abundance of $\psi_1$ is set by freeze-out and the abundance of $\psi_2$ by freeze-in.
 \item Both $\psi_1$ and $\psi_2$ thermalise, and their abundance is set by freeze-out.
\end{enumerate}
In region (A) the two abundances $Y_{1,2}$ evolve independently from one another, and therefore the isocontours of $\Omega$ have a simple structure. The minimum abundance in case of small couplings is given by the superWIMP contribution~\cite{Covi:1999ty,Feng:2003uy} from the late decay of frozen-out $\Sigma$.
In regions (B) and (C) the two yields $Y_1$ and $Y_2$ do not evolve independently, however their interplay is simple. Let us consider the region (B), where $\psi_2$ is thermalised, but not $\psi_1$. In this region, the production of $\psi_2$ is determined by the freeze-out through the annihilation $\psi_2\psi_2\rightarrow l^+ l^-$, which depends only on the coupling $g_2$ (for fixed masses), and is independent of the abundance of $\psi_1$. However, the abundance of the feebly coupled component $\psi_1$ can be affected by the abundance of the thermalised component $\psi_2$, if its coupling is sufficiently large.  More specifically, the abundance of $\psi_1$ is determined not only by two-body decays and scattering conversion of $\Sigma$ (as in the single flavor dark matter scenario), but also by the conversion process $\psi_2 l \to \psi_1 l$, which can dominate over the decay production if $g_2$ is sufficiently large. The same behaviour occurs in region (C) swapping $\psi_1$ by $\psi_2$. 

\begin{figure}[t!] 
  \begin{center}
    \includegraphics[width=0.45\textwidth]{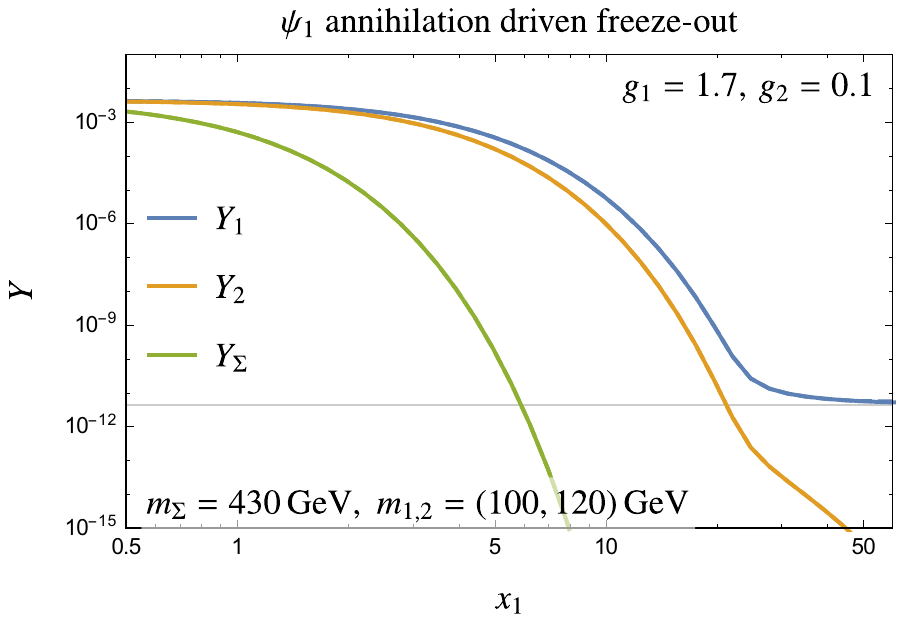}
    \includegraphics[width=0.45\textwidth]{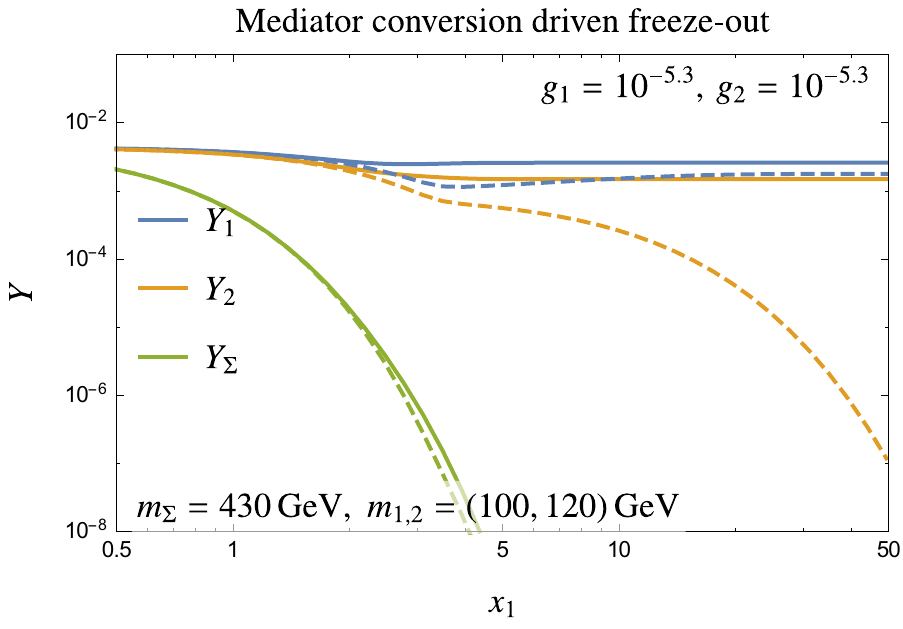}\\
    \includegraphics[width=0.45\textwidth]{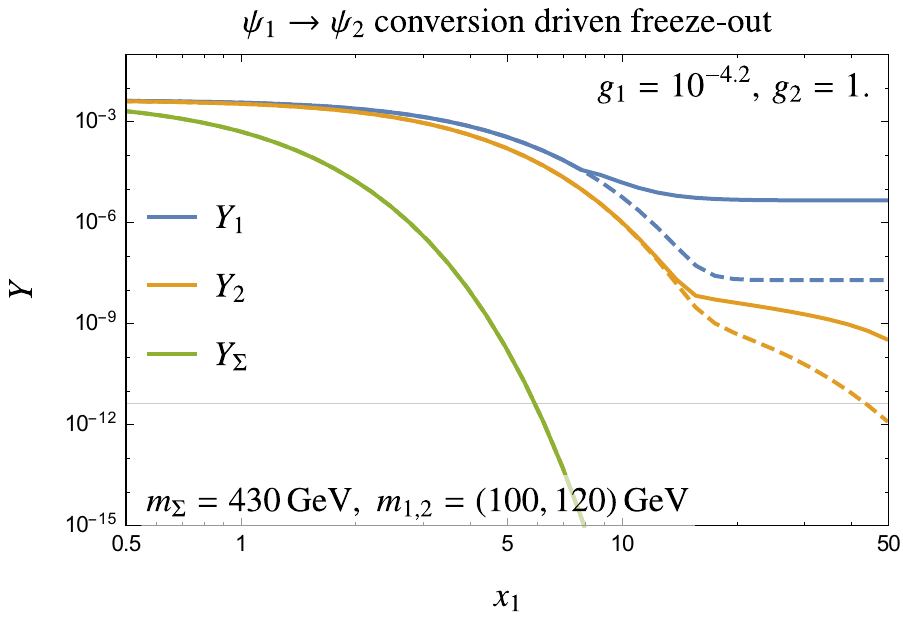}
    \includegraphics[width=0.45\textwidth]{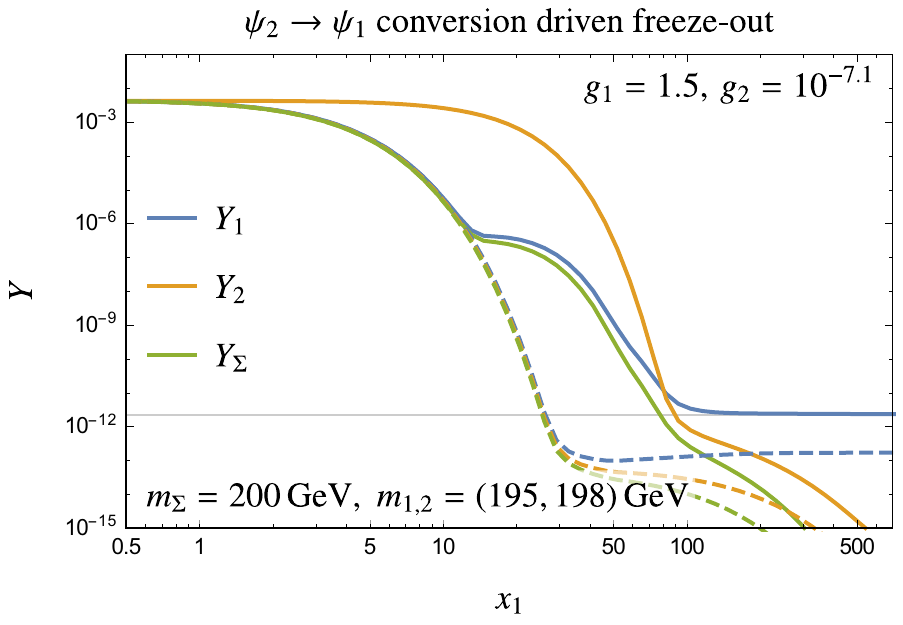}
  \end{center}
  \caption{
	Evolution with $x_1\equiv m_1/T$ of the abundances $Y_a$ of the different particle species of the dark sector $a=\psi_1$, $\psi_2$ and $\Sigma$, for various production regimes of region (D) in Fig.~\ref{fig:relicContours}. The dashed lines indicate the abundances assuming chemical equilibrium; the thin horizontal grey line indicates the yield corresponding to the observed dark matter abundance.
	}
	\label{fig:leptoAbundanceEvolution}
\end{figure}

The freeze-out region (D) exhibits several different regimes, indicated by numbers in Fig.~\ref{fig:relicContours}
\begin{enumerate}
 \item \emph{$\psi_1$ annihilation:} vertical contours at rather large $g_1$ for wide range of $g_2$. In this regime, the presence of $\psi_2$ is irrelevant to the relic abundance and the result is the same as in the standard single flavour WIMP case. The evolution of the abundances $Y_a(x)$ for this case is shown in Figure~\ref{fig:leptoAbundanceEvolution}, top-left panel.
 \item \emph{$\psi_2$ coannihilation:} horizontal contours at rather large $g_2$ for a wide range of smaller $g_1$.
 $\psi_2 \psi_2$-annihilation is the dominant dark sector depletion mechanism.
 \item \emph{$\Sigma$ mediator coannihilation:} For small values of $g_{1,2}$, neither $\psi_2\psi_2$ nor $\psi_1\psi_1$ annihilation is efficient. Instead, mediator coannihilation depletes the dark sector, which can be efficient if the mediator mass is not too far from $m_{1,2}$~\cite{Griest:1990kh}. The relic density here is determined by the SM gauge couplings and is roughly independent of $g_{1,2}$.
 \item \emph{Mediator conversion driven freeze-out:}
 The coannihilation plateau is bounded towards low couplings by rising relic abundances that are the result of $\psi_i$ dropping out of chemical equilibrium with $\Sigma$ before the dark matter abundance can be significantly depleted through $\Sigma$-coannihilations. This process has been described  in~\cite{Garny:2017rxs,DAgnolo:2017dbv,Junius:2019dci} and allows for thermalised dark matter with very small couplings, if the mass splitting between dark matter and the mediator is small. The evolution of the abundances $Y_a(x)$ is shown in  Figure~\ref{fig:leptoAbundanceEvolution}, top-right panel.
 \item \emph{$\psi_1 \to \psi_2$ conversion driven freeze-out:} Likewise, the $\psi_2$ coannihilation region is bounded towards small $g_1$ by $\psi_1$ dropping out of chemical equilibrium with $\psi_2$ before $\psi_2$-driven coannihilation can effectively deplete the $\psi_1$ abundance. This regime is specific to the two-flavour setup, and can be considered as an extension of the mediator conversion driven freeze-out. The evolution of the abundances $Y_a(x)$ is shown in  Figure~\ref{fig:leptoAbundanceEvolution}, bottom-left panel.
 \item \emph{$\psi_2 \to \psi_1$ conversion driven freeze-out:}
 At small $g_2$, $\psi_2$ becomes long lived and the dominant reaction depleting the relic abundance is $\psi_1 \psi_1$ annihilation. The relic abundance results from an interplay of $\psi_2 \to \psi_1$ conversion, out-of-equilibrium decay of $\psi_2$ and $\psi_1\psi_1$ annihilation.\footnote{
  If the decay of $\psi_2$ happens long after $\psi_1$ has frozen out and does not revive $\psi_1$ annihilation, the effect of $\psi_2$ is a simple additive contribution to the freeze-out $\Omega_1$ result (akin to the super-WIMP contribution in region A).
  This limit has been investigated in a 2-component WIMP scenario without scattering conversion by~\cite{Fairbairn:2008fb}. In general, conversion, decay and annihilation need to be considered together to determine the relic abundance in regime~6, as is evident from the evolution of the abundances $Y_a(x)$ shown in  Figure~\ref{fig:leptoAbundanceEvolution}, bottom-right panel.
 }
 This regime is also specific to the two-flavour setup and can result in the correct relic abundance for couplings $g_1$ much larger than in the single-flavour case.
\end{enumerate}
In regimes 1 to 3, chemical equilibrium between the dark sector particles holds during freeze-out. However, this is not the case for the regimes 4-6. There, the decoupling of conversion processes makes the depletion of dark matter abundance less effective, leading to larger relic densities. 

We also analyze for completeness scenarios with a compressed mass spectrum, $m_1 \sim m_2 \sim m_\Sigma$, where coannihilation processes are effective at depleting the dark sector. We show in Fig.~\ref{fig:relicContours}, middle panel, a representative example with $m_\Sigma=200$ GeV, $m_1=195$ GeV and $m_2=198$ GeV. One identifies the same regions (A), (B), (C) and (D) as in the generic benchmark, however with a larger allowed parameter space, due to coannihilations. Note that the relic abundance can be further reduced  when the quartic coupling $\lambda_{H\Sigma}$ is sizable, through processes such as $\Sigma\Sigma \to H^* \to \bar qq$.

Lastly, we have also explored the scenario where there is a large mass hierarchy between $\psi_1$ and $\psi_2$. The allowed parameter space is shown in  Fig.~\ref{fig:relicContours}, right panel, for the choice $m_\Sigma=430$ GeV, $m_1=1$ eV and $m_2=100$ GeV. 
In contrast to the previous two cases, $\psi_1$  does not contribute significantly to the dark matter density, even though its abundance is set  in large parts of the parameter space by relativistic freeze-out, and can potentially be large. Instead, $\psi_2$ is the dominant dark matter component. In this case, we only find one allowed region, which is bounded by the  requirement $g_2 \lesssim 2 \cdot 10^{-12} \sqrt{m_\Sigma/m_2}$ so that $\psi_2$ is not overproduced through freeze-in~\cite{Herms:2019mnu}, and by the requirement $\tau_{\psi_2}\gtrsim 4\times 10^{17}\,\mathrm{s}$, so that $\psi_2$ is sufficiently long-lived to contribute to the dark matter of the Universe today.

\section{Signatures of multi-flavour leptophilic Dark Matter}
\label{sec:leptoSignatures}
The single-flavored version of our leptophilic dark matter model  has been thoroughly studied in the literature (see eg.~\cite{Garny:2015wea,Belanger:2018sti}). It leads to a variety of signals in direct, indirect and collider search experiments, which are also present in our multi-flavor variant. In addition to these signals, the multi-flavor scenario leads to novel phenomena and novel signatures that increase the discovery potential of the multi-flavored scenario compared to the single-flavored one. 

\begin{figure}[t!]
	\begin{center}
	 \includegraphics[width=0.32\textwidth]{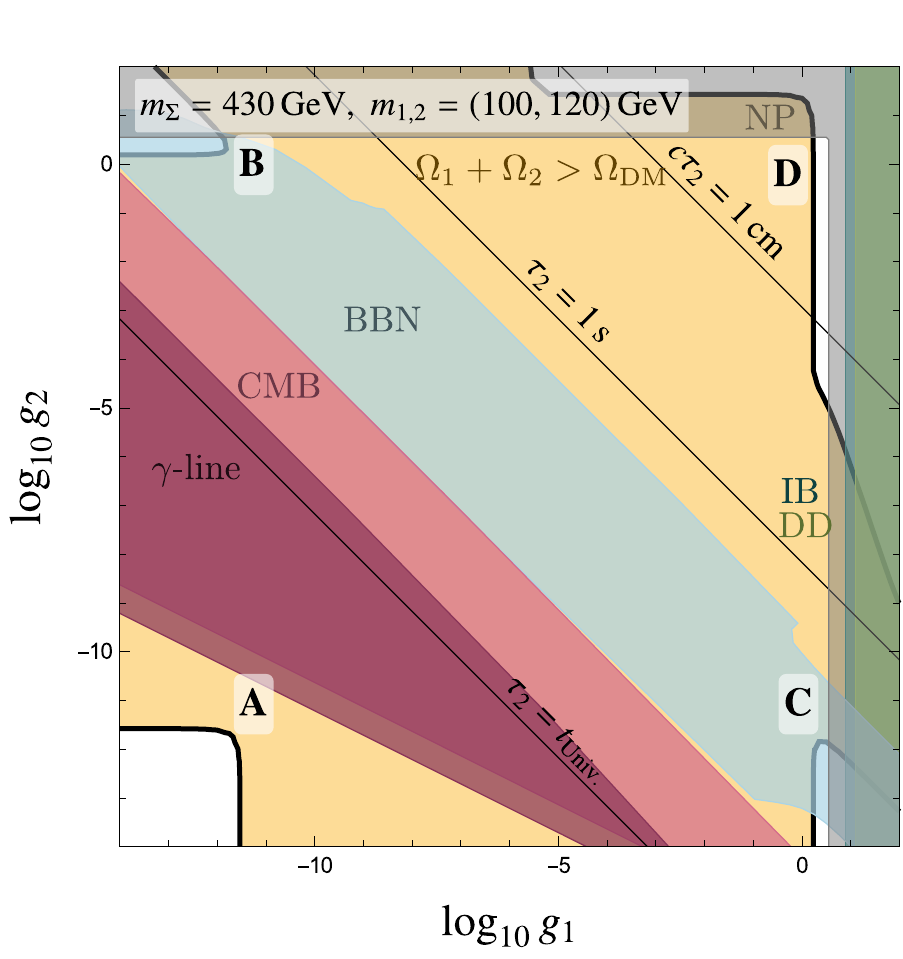}
	 \includegraphics[width=0.32\textwidth]{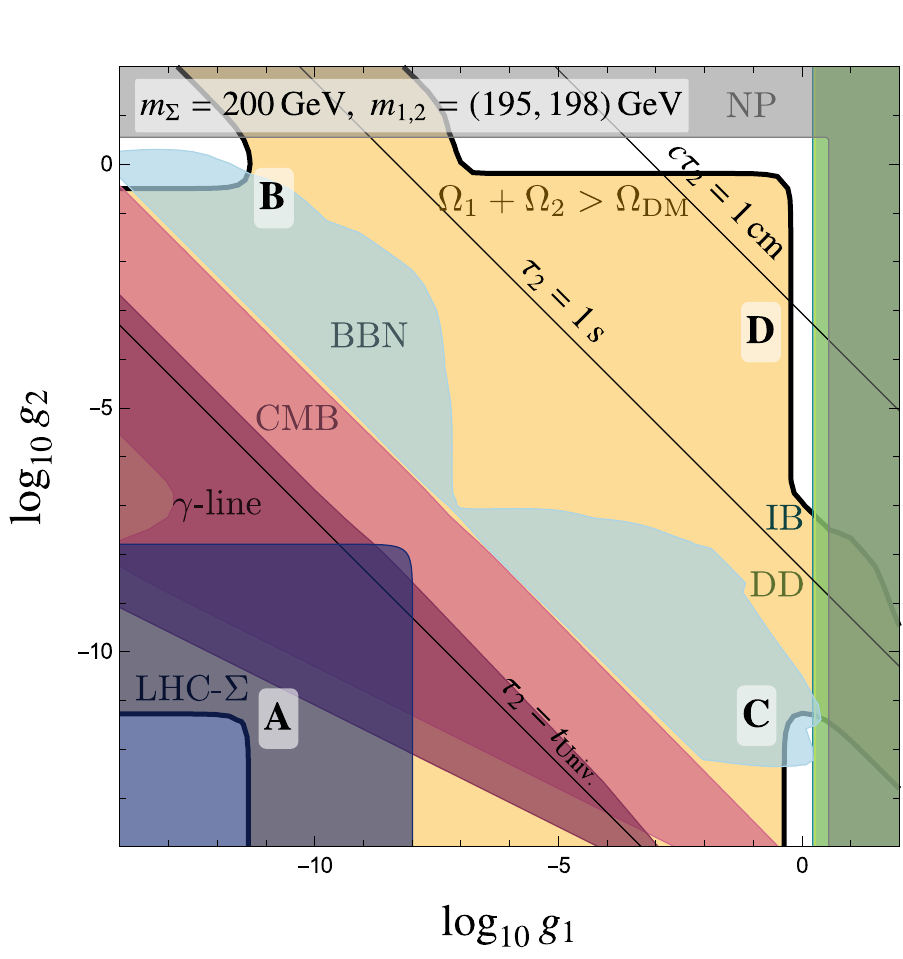}
	 \includegraphics[width=0.32\textwidth]{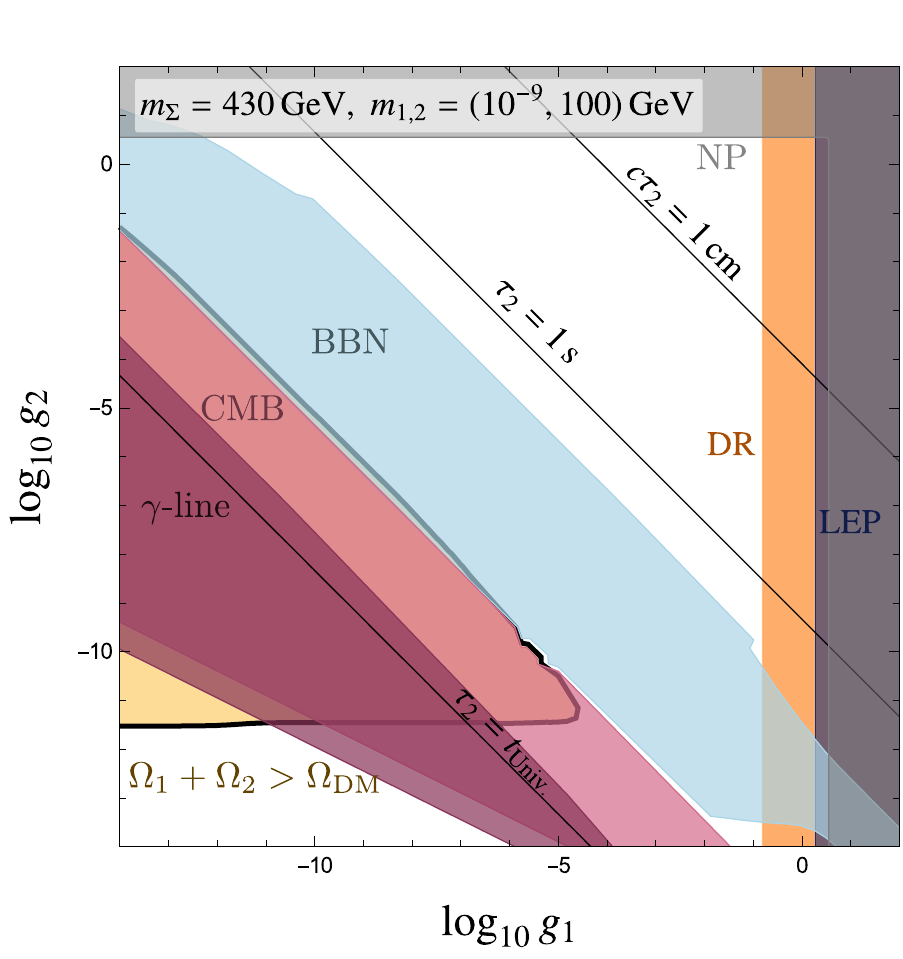}
	\end{center}
	\caption{Isocontours of the lifetime (or decay length) and constraints on the benchmark scenarios of Fig.\ref{fig:relicContours} of our multi-flavor leptophilic dark matter model
	Concretely we show the constraints from non-perturbativity (NP), overabundance ($\Omega_1+\Omega_2>\Omega_\mathrm{DM}$), Big Bang Nucleosynthesis (BBN), Cosmic Microwave Background (CMB) , $\gamma$-rays produced in the decay $\psi_2\rightarrow\psi_1\gamma$ ($\gamma$-line), $\gamma$-rays produced in the internal bremsstrahlung annihilation process $\psi_1\psi_1\rightarrow l^+l^-\gamma$ (IB), direct detection through the $\psi_1$-anapole moment (DD), LHC searches for anomalous charged tracks left by the mediator  $\Sigma$ (LHC-$\Sigma$), dark radiation (DR) and monophoton searches at LEP (LEP). The allowed regions are shown in white.
	}
	\label{fig:leptoDensityWithLimits}
\end{figure}

One of the most distinct characteristics of the multi-flavor dark matter scenario is the decay of the heavier generation $\psi_2$ into the lighter (stable) generation $\psi_1$ along with Standard Model particles. The dominant decay channel is $\psi_2\rightarrow \psi_1 l^+ l^-$ with rate given in Eq.~(\ref{eqn:fermionDecayRateThreeBody}). 
Fig.~\ref{fig:leptoDensityWithLimits} shows the isocontours of the $\psi_2$ lifetime for the three exemplary scenarios presented in Section~\ref{sec:fermionWIMP}. The white regions are defined by requiring that the sum of the densities of $\psi_1$ and $\psi_2$ does not exceed the dark matter density at the time of recombination. In the generic benchmark and the near degenerate benchmark, there are four distinct regions (A), (B), (C), (D) characterized by whether $\psi_1$ or $\psi_2$  freeze-in or freeze-out from the thermal plasma. In the very hierarchical benchmark, the allowed parameter space opens-up considerably, and these four regions are not separated by dark matter overabundance.

As apparent from the plot, in region (D) $\psi_2$ has a lifetime shorter than $\sim 1$ ms; in regions (B) and (C) $\psi_2$ is long lived, although no longer present in the Universe; and in region (A), $\psi_2$ is cosmologically long-lived and contributes to the dark matter density of the Universe today. Let us discuss in what follows the possible signatures arising in the decay of $\psi_2$ and the corresponding probes of the allowed regions of our multiflavor dark matter scenario in direct and indirect searches, in collider experiments, or through their impact on the early Universe.

If $\tau_{\psi_2}\lesssim 10^{-12}\second$, $\psi_2$ decays promptly inside the detector and is constrained by the search for hard leptons plus missing transverse energy, or in the degenerate case by the search for initial state radiation plus missing transverse energy.
If  $10^{-12}\second \lesssim \tau_{\psi_2} \lesssim 10^{-5}\second$, $\psi_2$ produces a displaced signature also in the detector, with (proper) decay lengths $1 \,\mathrm{mm} \lesssim c\tau_{\psi_2} \lesssim 1 \,\mathrm{km}$, which could be observed in future experiments. The search for these phenomena allows in principle to probe region (D) of the allowed parameter space. However, although searches for these signatures at the LHC exist (e.g.~\cite{CMS:2014hka,Aad:2019tcc}), signals in our scenario are in part preempted by the softness of the produced leptons and constraints on the production of $\Sigma$.
At lepton colliders, $\psi_i$ can be produced directly, leading to constraints from LEP on $g_i$ for sufficiently light $m_i$~\cite{Fox:2011fx}. Future lepton colliders may extend the reach in $m_i$~\cite{Horigome:2021qof}, widening the prospect of probing $\psi_2$ decay.

For longer lifetimes  $\psi_2$ escapes undetected, although it could leave an imprint in cosmological and astronomical observations. For $c\tau_{\psi_2}\gtrsim 10 \,\mathrm{m} \left(m_1/100\GeV\right)^{-2}$, $\psi_1 \psi_1$ annihilations leading to the observed relic abundance are no longer in equilibrium when $\psi_2$ decays in the early Universe, modifying the dark matter relic abundance. This is one of the most salient differences of our multi-flavored dark matter scenarios compared to the single-flavored limit. 

\begin{figure}[t!]
	\begin{center}
	 \includegraphics[width=0.43\textwidth]{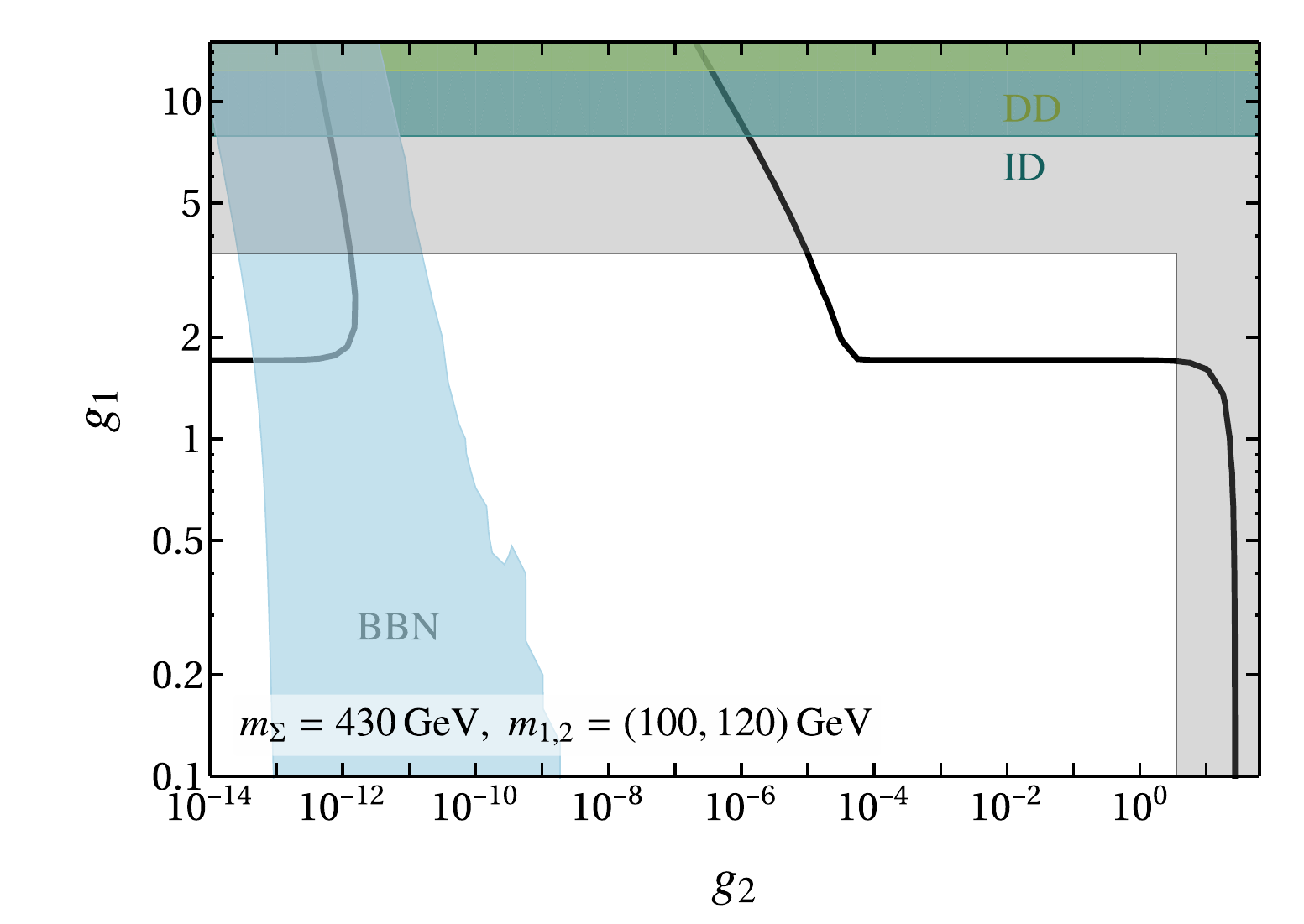}
	 \includegraphics[width=0.43\textwidth]{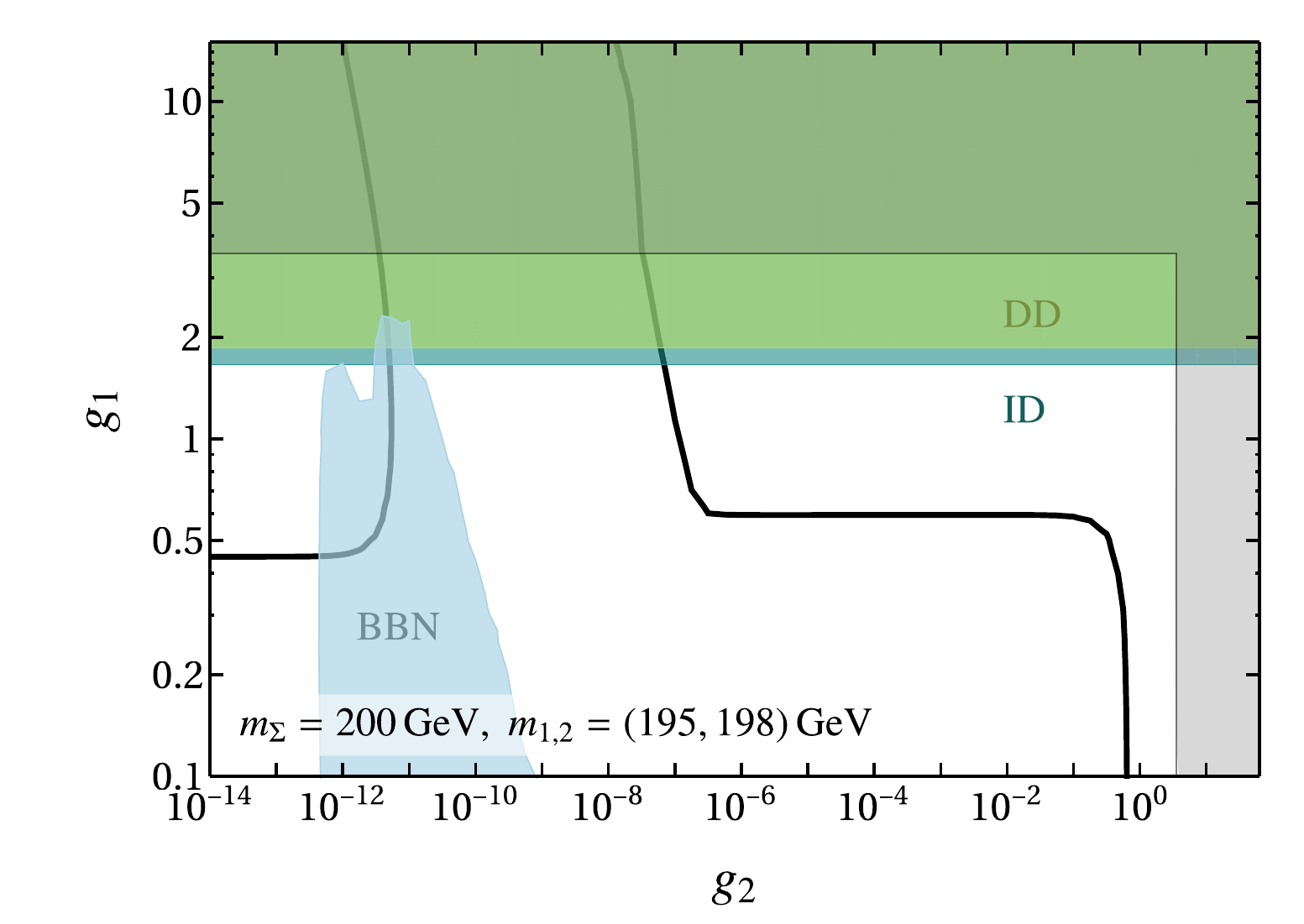}
	\end{center}
	\caption{Values of the Yukawa couplings $g_1$ and $g_2$ leading to the observed dark matter abundance in the generic benchmark (left panel) and in the near-degenerate benchmark (right panel) of our multiflavor dark matter scenario. The light green region is excluded by direct detection experiments, the dark green region by indirect detection experiments, and the light blue region by Big Bang Nucleosynthesis; in the grey region perturbation theory loses validity.
	}
	\label{fig:annihilationEnhancement}
\end{figure}

This is illustrated in 
Figure~\ref{fig:annihilationEnhancement}, which shows the values of the couplings $g_1$ and $g_2$ that lead to the correct relic abundance (black line) for the generic benchmark (left panel) and for the nearly-degenerate benchmark (right panel). In the region above the line, $\psi_1$ does not account for all the dark matter of the Universe, but is not necessarily ruled out. We also show in grey the regions where our perturbative approach loses validity.  Again, these regions are not necessarily ruled out, but are removed for consistency of the calculation. The single-flavored limit corresponds to $g_2=0$, in which $\psi_2$ completely decouples. One sees from the Figure that the presence of $\psi_2$ can lead to an enhancement of the value of $g_1$ necessary to obtain the same dark matter abundance. For small $g_2$, this is due to the $\psi_1$ production due to 
the decay of frozen-in $\psi_2$, and for larger $g_2$ due to modifications of the freeze-out process through the continuous conversion and decay of $\psi_2$.
At very large $g_2$, coannihilation processes reduce the value of $g_1$ required to reproduce the observed relic abundance.
We also show in the plot the current limits on $g_1$ for those benchmark scenarios from direct (light green) and from indirect detection experiments (dark green). Concretely, for the direct detection limits, we have recast the limits derived in \cite{Kang:2018oej} from the non-observation of nuclear recoils at the \textsc{Xenon1T} experiment~\cite{Aprile:2018dbl} induced by the dark matter anapole moment.\footnote{For coupling to electrons, dark matter-electron scattering could induce atomic ionizations. However, current limits from the \textsc{Xenon1T} experiment~\cite{Aprile:2019xxb} do not constrain couplings compatible with our perturbativity requirement.} These limits are particularly relevant for the very degenerate regime~\cite{Kopp:2014tsa}, as illustrated in the right plot. For the indirect detection limits, we have used the results derived in \cite{Bringmann:2012vr} from the non-observation of signals from the two-to-three annihilation $\psi_1 \psi_1 \to l^+ l^- \gamma$ at the center of the Milky Way (the rate of the two-to-two annihilation  $\psi_1 \psi_1 \to l^+ l^-$ is negligibly small due to a $p$-wave suppressed cross-section). This process has been studied {\it e.g.} in~\cite{Garny:2011ii,Bringmann:2012vr,Kopp:2014tsa,Garny:2013ama,Garny:2015wea} and leads to a distinct spectral feature in the gamma ray spectrum. While experiments are yet far from probing the single-flavour scenario for our adopted benchmark scenarios, the presence of the heavier flavour in the early Universe enhances the potential of observing signals in the future.

For lifetimes $\tau_{\psi_2} \gtrsim 1\second$, $\psi_2$ can directly affect well-established cosmological observables.
The exotic energy injection from $\psi_2$ decay can modify the primordial abundances of light elements in Big Bang Nucleosynthesis~\cite{Kawasaki:2020qxm} and lead to spectral distortions of the CMB~\cite{Poulin:2016anj}. For $\tau_{\psi_2} \gtrsim 10^{12}\second$, $\psi_2$ is long lived enough to contribute to the dark matter relic density measured by the CMB, and its decay products can modify the ionization history between recombination and reionisation, affecting the CMB angular power spectra \cite{Poulin:2016anj,Slatyer:2016qyl}. Both BBN and the CMB are sensitive probes of regions (B) and (C) of the parameter space. 

\begin{figure}[t!]
	\begin{center}
	    \includegraphics[width=0.55\textwidth]{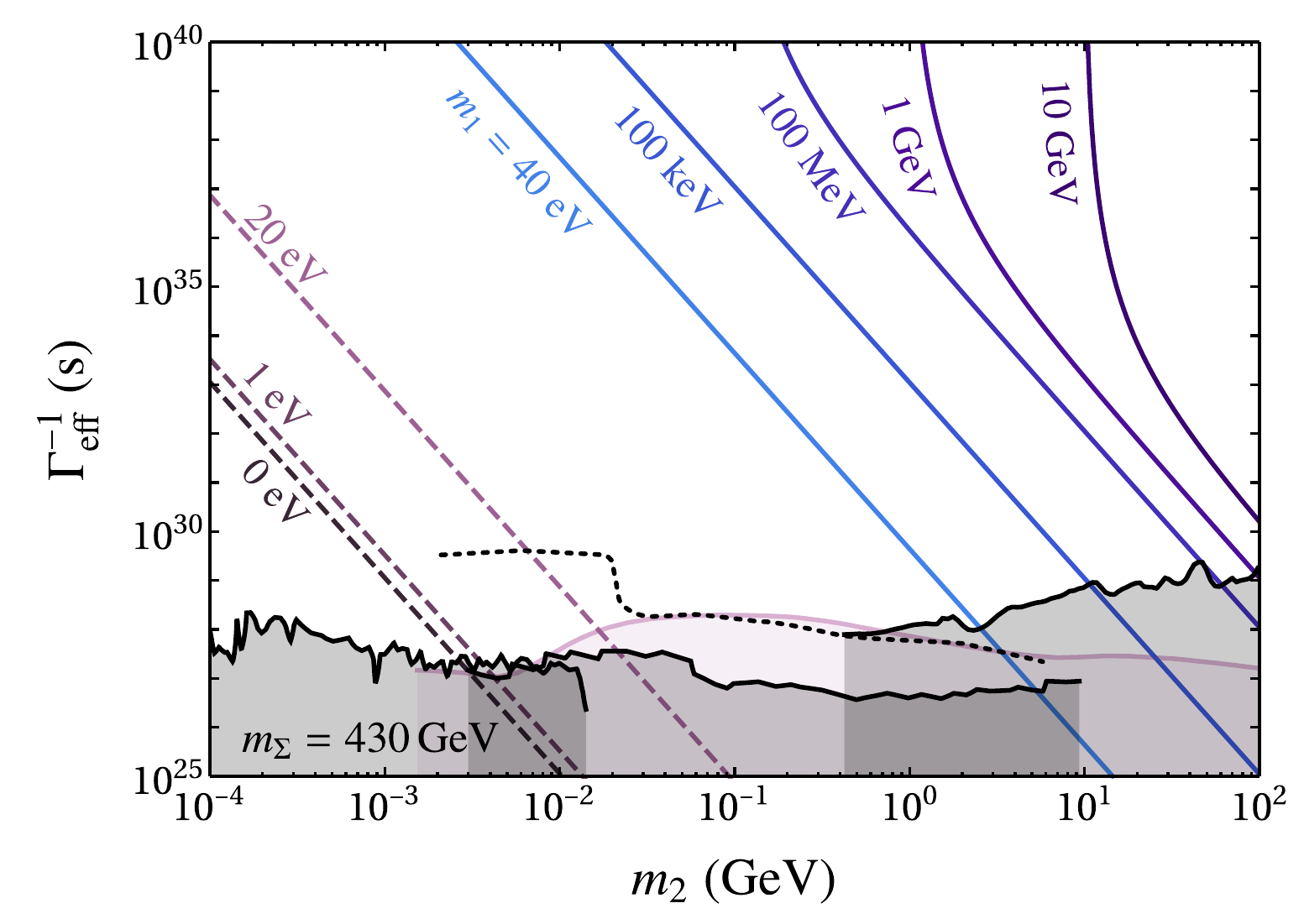}
	\end{center}
	\caption{
	 Lower limit on the inverse decay rate for $\psi_2\rightarrow\psi_1\gamma$ in our multi-flavor leptophilic dark matter scenario, as a function of the mass of the decaying FIMP component $m_2$ for different values of the mass of the stable FIMP component $m_1$. The FIMP components are assumed to have a Yukawa coupling to a heavy scalar $\Sigma$, with mass fixed to $430\GeV$, and to a right-handed electron. The grey regions correspond to the lower limit on the rate from the non-observation of a statistically significant sharp feature in the isotropic diffuse photon flux, and the pink regions to the recast limit on the rate from the non-observation of signatures of the decay $\psi_2\rightarrow \psi_1 e^+ e^-$ in CMB data. For the masses indicated by solid lines (dashed lines), $\psi_1$ constitutes a subdominant component of dark matter (dark radiation). The black dotted line indicates the reach of the proposed e-\textsc{Astrogam} mission concept. 
	}
	\label{fig:fermionDecayResult}
\end{figure}

Finally, if  $\tau_{\psi_2} \gtrsim 10^{18}\second \sim t_\mathrm{Univ}$, $\psi_2$ is cosmologically long lived and can contribute to the dark matter density today. Yet, the Standard Model particles produced in the decay could be detected in experiments, thus providing a unique probe of the otherwise elusive region (A) of the parameter space.
Figure~\ref{fig:fermionDecayResult} shows the maximal gamma ray signal from the process $\psi_2\rightarrow\psi_1\gamma$ under the requirement that the relic density does not exceed the measured dark matter density.
Blue lines correspond to the maximal signal $\Gamma_\mathrm{eff} \equiv \Gamma \Omega_2/\Omega_\mathrm{DM}$ obtainable for the case where neither of the dark matter fermions thermalize in the early Universe, using the results of~\cite{Herms:2019mnu}.
These are compared to the limits on dark matter decay into monochromatic gamma rays calculated in~\cite{Essig:2013goa} based on measurements by the \textsc{Comptel}~\cite{1999ApL&C..39..193W}, \textsc{Egret}~\cite{Strong:2004de} and \textsc{Fermi-LAT}~\cite{Ackermann:2014usa} instruments, as well as the dedicated line-search in the \textsc{Integral} data~\cite{Boyarsky:2007ge} and by the \textsc{Fermi-LAT} collaboration~\cite{Ackermann:2015lka}.
We find that the multi-component fermionic FIMP scenario can be probed by current experiments, if the stable DM component is much lighter than the decaying one, $m_1 \ll m_2$, concretely when
$m_2\gtrsim 3 \GeV$ for $m_1=40 \eV$, when 
$m_2\gtrsim 40 \GeV$ for $m_1=100 \MeV$, or when
$m_2\gtrsim 90\GeV$ for $m_1=1 \GeV$.

The electrons and positrons produced in the decay $\psi_2 \to \psi_1 e^+e^-$ may have observable consequences and thereby provide complementary probes of our scenario. Concretely, the charged particles injected into the SM plasma could affect the reionization history of our Universe. We show in Fig.~\ref{fig:fermionDecayResult} as a pink region the part of parameter space where the observation of the gamma-ray line is precluded by the reionization constraints derived in~\cite{Slatyer:2016qyl}.
We also note that at mass differences $m_2-m_1 \gtrsim 10\GeV$, searches for the associated spectral feature in the spectrum of cosmic ray positrons are promising (see e.g.~\cite{Bergstrom:2013jra,Ibarra:2013zia} for existing searches for positron spectral features) and complementary to the gamma ray signatures we focus on here.

In the case where both DM flavours are FIMPs, gamma ray signals at $E_\gamma \lesssim 1\GeV$ are inaccessible to current instruments, as the large couplings necessary to produce an observable signal would violate DM overabundance (larger $\Gamma$ requires larger couplings, leading to larger freeze-in yield) or free-streaming limits ($m_\mathrm{FIMP}\lesssim 10 \keV$ are only allowed to make up a small fraction of $\Omega_\mathrm{DM}$, constraining $g_1$ in the FIMP case).
However, for very light $m_1$, $\psi_1$ could thermalize in the early Universe, allowing for large coupling $g_1$ without overproducing dark matter. This can lead to large decay signals down to MeV energies (dashed lines in Fig.~\ref{fig:fermionDecayResult}).
The maximal signal in this case is determined by limits on the allowed fraction of non-cold dark matter or dark radiation in the form of relativistically-decoupling $\psi_1$.
We use free-streaming limits on the fraction of non-cold dark matter from~\cite{Diamanti:2017xfo} to find a lower limit on the freeze-out temperature $T_\mathrm{dec}$ of $\psi_1$, depending on $m_1$. This fixes $m_1 \leq 20\eV$ for relativistically-decoupling $\psi_1$ and leads to an upper limit on $g_1$ corresponding to a maximal dark matter decay signal.

In the case where the heavier DM fermion is produced by freeze-in while the lighter one freezes-out while relativistic, gamma ray signals down to $E_\gamma \sim \MeV$ are possible, where future gamma-ray telescopes like the proposed \textsc{Amego}~\cite{McEnery:2019tcm} or e-\textsc{Astrogam}~\cite{Bartels:2017dpb,DeAngelis:2017gra} mission concepts could improve the sensitivity by 1-2 orders of magnitude.

\section{Conclusions}
\label{sec:conclusion}
In this work, we have studied possible effects of the existence of multiple dark matter generations in Nature, akin to the existence of multiple fermion generations in the Standard Model. We have concentrated for concreteness on a scenario with two generations of Majorana fermion dark matter candidates, $\psi_1$ and $\psi_2$, that couple via a Yukawa coupling to a right-handed charged lepton and a scalar mediator, $\Sigma$. While the quantitative results are very model dependent, the rationale for the analysis and some qualitative conclusions are completely general.

We have identified several processes which are relevant for setting the dark matter relic abundance depending on the region of the parameter space of the model: $\psi_1$ annihilation, $\psi_2$ annihilation, $\Sigma$ mediator coannihilation, mediator conversion driven freeze-out, and $\psi_1\rightarrow \psi_2$ or $\psi_2\rightarrow \psi_1$ conversion driven freeze-out, as well as freeze-in. We have found that in some instances the values of the couplings of the lighter (stable) dark matter component leading to the observed dark matter abundance can be much larger than the one expected in the single flavored scenario. This opens the possibility of an enhancement of the signal strength in dark matter search experiments, without invoking astrophysical boost factors.

For the regions allowed by current determinations of the dark matter abundance, we have investigated the possible implications of the decay of the heavier generation into the lighter in the early Universe and for dark matter search experiments. The model leads to a wealth of possible signals which probe different parts of the allowed parameter space, in particular parts which would be hard or impossible to probe in a single-flavor dark matter model.

\section*{Acknowledgments}
This work has been partially funded by the Collaborative Research Center SFB1258 and by the Deutsche Forschungsgemeinschaft (DFG, German Research Foundation) under Germany's Excellence Strategy – EXC-2094 – 390783311.

\normalem 
\bibliographystyle{JHEP}
\bibliography{refs}
\end{document}